\documentclass[conference,10pt]{IEEEtran}
\usepackage{stmaryrd}

\usepackage[dvips]{graphicx}
\usepackage[english]{babel}
\usepackage[latin1]{inputenc}
\usepackage{algorithm2e}
\usepackage{amsmath}

\usepackage{amssymb}
\usepackage{url}

\usepackage{cite} 
\usepackage{stfloats}  

\newcommand{\ud}{\mathrm{d}}
\newcommand{\Var}{\mathrm{Var}}
\graphicspath{{Graphs/}}
\SetKw{KwFn}{function}
\DeclareMathOperator*{\argmin}{argmin}
\IEEEoverridecommandlockouts

\begin{document}

\title{Dimensionally Distributed Learning \\ Models and Algorithm\thanks{This research was supported in part by the U.S.
National Science Foundation under Grants ANI-03-38807 and
CNS-06-25637, and by the Office of Naval Research under contract
numbers W911NF-07-1-0185 and N0014-07-1-0555.}}

\author{
\authorblockN{Haipeng Zheng}
\authorblockA{Department of Electrical Engineering\\
Princeton University\\
Princeton, NJ, U.S.A.\\
Email: haipengz@princeton.edu} \and
\authorblockN{Sanjeev R. Kulkarni}
\authorblockA{Department of Electrical Engineering\\
Princeton University\\
Princeton, NJ, U.S.A.\\
Email: kulkarni@princeton.edu} \and
\authorblockN{H. Vincent Poor}
\authorblockA{Department of Electrical Engineering\\
Princeton University\\
Princeton, NJ, U.S.A.\\
Email: poor@princeton.edu}}

\maketitle

\selectlanguage{english}

\begin{abstract}
This paper introduces a framework for regression with dimensionally
distributed data with a fusion center. A cooperative learning
algorithm, the iterative conditional expectation algorithm (ICEA),
is designed within this framework. The algorithm can effectively
discover linear combinations of individual estimators trained by
each agent without transferring and storing large amount of data
amongst the agents and the fusion center. The convergence of ICEA is
explored. Specifically, for a two agent system, each complete round
of ICEA is guaranteed to be a non-expansive map on the function
space of each agent. The advantages and limitations of ICEA are also
discussed for data sets with various distributions and various
hidden rules. Moreover, several techniques are also designed to
leverage the algorithm to effectively learn more complex hidden
rules that are not linearly decomposable.
\end{abstract}


%
\IEEEpeerreviewmaketitle
\section{Introduction}
Distributed learning is a field that generalizes classical machine
learning algorithms. Instead of having full access to all the data
and being capable of central computation, in the framework of
distributed learning, there are a number of agents that have access
to only part of the data. And the agents (perhaps with a fusion
center) are capable of exchanging certain types of information among
one another. Usually, due to privacy concerns, limited bandwidth and
limited power, the content and amount of information shared are
restricted. Research in distributed learning seeks effective
learning algorithms and theoretical limits within such constraints.

In terms of data structures, two types of distributed learning
problems are: homogeneous data and heterogeneous data (or
horizontally distributed / instance distributed data and vertically
distributed / dimensionally distributed data). In terms of the
organization of distributed learning systems, there are also
basically two types: systems with a fusion center and systems
without a fusion center. In \cite{ref1}, \cite{ref3} two important
types of models, instance distributed learning with and without a
fusion center, are discussed and several practical algorithms are
provided. The relationship between the information transmitted
amongst individual agents and the fusion center and the ensemble
learning ability are also discussed. There has also been some
research on distributed learning with dimensionally distributed
data, such as in \cite{ref2}. However, the approach in \cite{ref2}
is not cooperative - individual agents first optimize their own
estimator, and, given these estimators, the fusion center then
constructs an optimal linear combination of them. In this paper,
however, we will concentrate on a cooperative training algorithm, in
which the fusion center coordinates the individual agents to
optimize the ensemble estimator.

It is also worth pointing out the connection between dimensionally
distributed learning and boosting for regression, which was first
introduced in \cite{ref6} and developed in many other works such as
\cite{ref7}. The algorithm developed in this paper can be viewed as
an $L_2$-regression boosting algorithm with extra constraints on the
space from which the weak hypothesis can be selected. This
perspective can bring insights of boosting to the problem of
distributed regression.

\section{Description of the Problem}
In this paper, we discuss the problem of estimation (or regression)
with dimensionally distributed data and a fusion center. The problem
is specified as follows.

There are $M$ independent variables (or features) $X_1,\cdots,X_M$
and one dependent variable $Y$. The complete data set is composed of
\begin{displaymath}
\{(x_{i1},x_{i2}, \cdots, x_{iM},y_{i})\}_{i=1}^{n}
\end{displaymath}
where $n$ is the number of instances, $x_{ij}\in \mathbb{R}$ is the
$i$-th instance of $X_j$, and $y_{i}\in \mathbb{R}$ is the $i$-th
instance of $Y$.

We also assume that there exists a hidden deterministic function (or
rule, or hypothesis)$$\phi: \mathbb{R}^M \rightarrow
\mathbb{R}$$such that$$y_i = \phi(x_{i1}, x_{i2},\cdots ,x_{iM}) +
w_i$$ where $\{w_i\}_{i=1}^{n}$ is an independently drawn sample
from a zero-mean random variable $W$ that is independent of
$X_1,\cdots,X_M$ and $Y$.

Suppose there are $D$ agents, each of which has only limited access
to certain features. Define $F_j (j=1,\cdots,D)$ to be the set of
features accessible by agent $j$, and define $F = \cup _{j=1}^D F_j$
so that $|F|=M$.

In order to concentrate on the ``distributed part" of the problem,
we assume that each agent is capable, given enough data, of learning
the optimal minimum-mean-square-error (MMSE) estimator based on
limited access to the features. More specifically, we assume that
agent $j$ can solve the optimization problem (given enough data)
$$
\min_{g_{j}(\{X_t\}_{t \in F_j})} \mathbb{E}\left[ \left(
\zeta(\{X_t\}_{t \in F}) - g_{j}(\{X_t\}_{t \in F_j}) + W\right)^2
\right]
$$
where $\zeta$ can be any $M$ dimensional function satisfying some
regularity conditions. Due to the independence and unbiasedness of
the noise, the above optimization problem can be simplified to
$$
\min_{g_{j}(\{X_t\}_{t \in F_j})} \mathbb{E}\left[ \left(
\zeta(\{X_t\}_{t \in F}) - g_{j}(\{X_t\}_{t \in F_j})\right)^2
\right]
$$
The solution to the optimization problem above is
$$g_{j}(x_{F_j}) = \mathbb{E}[\zeta(X_{F})|
X_{F_j}=x_{F_j}]$$ where, for simplicity, we use $x_{F_j}$ to
represent $\{x_t\}_{t\in F_j}$; that is, we assume that each agent
is capable of estimating the conditional expectation of a function
on $F$ given the several dimensions comprising $F_j$, and based on
enough data.

Under this model, one way to deal with the distributed estimation
problem is another optimization problem formulated as follows:
$$
\min_{\rho(g_1, g_2,\cdots , g_D)}\mathbb{E}\left[\left( \phi(X_F) -
\rho\left(g_1(X_{F_1}),\cdots , g_D(X_{F_D})\right) \right)^2\right]
$$
where the functions $g_i,i=1,\cdots,D$ are fixed and given by the
agents. The optimization problem above is intractable in its full
generality, and this non-cooperative training approach does not take
full advantage of the communication between individual agents and
the fusion center (because it uses only one-way agent-to-center
communication).

However, if we restrict the function $\rho$ to be of the additive
form
$$\rho(g_1, g_2,\cdots,g_D)=g_1+g_2+\cdots+g_D,$$ and optimize over $g_j,j=1,\cdots,D$, i.e. we change this problem into a
simplified version
$$
\min_{g_1, g_2,\cdots , g_D}\mathbb{E}\left[\left( \phi(X_F) -
\left(g_1(X_{F_1})+\cdots +g_D(X_{F_D})\right) \right)^2\right],
$$
we then change a two-step optimization problem (first individual
agents optimize their own estimators, then the fusion center
optimizes the ensemble) to a one-step cooperative optimization
problem (the agents, with the coordination of the fusion center,
optimize the sum of their estimators cooperatively). So we can seek
an algorithm through which the agents can cooperatively solve the
above problem.

\section{Communication and Memory Restrictions}
We assume that each agent can store all the data instances of its
accessible features, i.e. agent $j$ has access to data
$\{x_{it}\}_{i=1}^n, \forall t \in F_j$. We also assume that the
fusion center can store $\{y_i\}_{i=1}^n$, which is equivalent to a
one-dimensional data set. The agents and fusion center also have an
additional one-dimensional memory (which can store all the instances
of the dependent variable or one dimension of the features) for
computation only, and there is no additional space beyond their own
allocation.

We further assume that the fusion center has two-way communication
with all the agents. To be more specific, each agent can read and
write on the one-dimensional data stored in the fusion center.

\begin{figure}[!hbtp] 
  \centering
  \includegraphics[scale = 1.0]{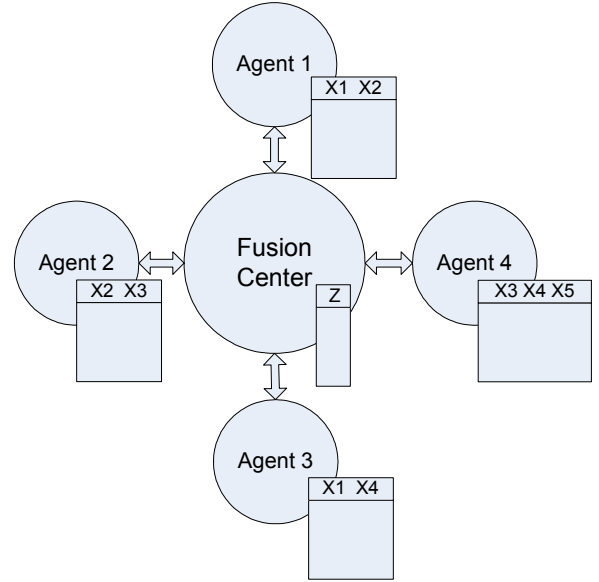}
  \caption{Structure of a dimensionally distributed learning system for regression.}
  \label{struct}
\end{figure}

Moreover, as noted above, we also require each agent to be capable
of finding the ideal MMSE estimator (within a certain function space
$\mathcal {F}$) based on its accessible data. Fig.\ref{struct} is an
illustration of the structure of a typical dimensionally distributed
learning system.

\section{Iterative Conditional Expectation Algorithm}
\subsection{Basic Idea}

Motivated by the \emph{backfitting algorithm} for additive models in
\cite{ref5}, we propose the iterative conditional expectation
algorithm (ICEA). The basic idea of this algorithm is very simple:
First, agent 1 asks for the value of $\phi(x_F)$ for all the data
instances from the fusion center, makes an estimate based on
features in $F_1$ and thereby obtains $g_1(x_{F_1})$. Of course,
$g_1(x_{F_1})$ cannot fully represent the true function $\phi(x_F)$
because it lies in a much smaller function space.

Then, agent 1 sends back its estimate for all the data instances to
the fusion center, and, the fusion center stores the residual
$\phi(x_F) - g_1(x_{F_1})$ for all the data instances. Then, agent 2
asks for the value of $\phi(x_F) - g_1(x_{F_1})$ from the fusion
center, makes an estimate based on features in $F_2$ and thereby
obtains $g_2(x_{F_2})$. This time, $g_1(x_{F_1}) + g_2(x_{F_2})$ is
a better approximation of the true hidden rule $\phi(x_F)$.

This process is continued for all agents. When the process
eventually returns to agent 1, it then asks the fusion center for
the value of $\phi(x_F)-\sum_{j=1}^D g_j(x_{F_j})$, thereby obtains
$\Delta g_1(x_{F_1})$, and stores $(g_1+\Delta g_1)(x_{F_1})$ as the
updated version of $g_1(x_{F_1})$. Then agent 1 sends that value of
$\Delta g_1(x_{F_1})$ for all the instances to the fusion center,
and the fusion center obtains the updated version of
$\phi(x_F)-\sum_{j=1}^D g_j(x_{F_j})$. This continues to agent 2,
and so on.

After a few rounds of iteration, the algorithm will converge to a
limit (we will show this below, under some further conditions). And
the sum of the limit of the functions, i.e. $$\sum_{j=1}^D
g_j(x_{F_j})$$ is the best linearly decomposed approximation of
$\phi(x_F)$ in terms of MMSE.

\subsection{ICEA in Detail}
The following is a more precise description of the above algorithm
(in terms of actual data instead of an evolution of ideal
functions):

\begin{procedure}[H]
\SetLine $ g_j(x_{F_j})\leftarrow 0, \forall j\in \{1,\cdots,D\}$\;
$z_i \leftarrow y_i, \forall i\in\{1,\cdots,n\}$\; $err_{new}
\leftarrow \frac{1}{n} \sum_{i=1}^n z_i^2$\; $err_{old} \leftarrow
0$\; \While{$|err_{old} - err_{new}|
> \epsilon$}{ \For{$j$ from $1$ to $D$} {
$\Delta g_j(x_{F_j}) \leftarrow $TRAIN$( \{ z_i, \{x_{it}\}_{t\in
F_j} \}_{i=1}^n)$\; $g_j(x_{F_j}) \leftarrow g_j(x_{F_j}) + \Delta
g_j(x_{F_j})$\; $z_i \leftarrow z_i - \Delta g_j(\{x_{it}\}_{t\in
F_j}), \forall i$\;
 }
 $err_{old} \leftarrow err_{new}$\;
 $err_{new} \leftarrow \frac{1}{n}
\sum_{i=1}^n z_i^2$\;
 }

\end{procedure}

\begin{function}[H]

 \KwFn TRAIN($\{y_i, \{x_{it}\}_{t\in F}\}_{i=1}^n $) \Return{$g(x_F)$}\;
 $g(x_F) =  \argmin_{g\in\mathcal {F}} \sum_{i=1}^n \left( y_i-g(\{x_{it}\}_{t\in F})\right)^2
 $

\end{function}

Actually, the $\Delta g_j(x_{F_j}) \leftarrow $TRAIN$ ( \{ z_i,
\{x_{it}\}_{t\in F_j} \}_{i=1}^n)$ step, given enough data, is
essentially computing
$$
\Delta g_j(x_{F_j}) \leftarrow \frac{\int_{F \setminus F_j}
\zeta(x_F) f_j(x_{F_j}) \ud \mu}{\int_{F \setminus F_j} f_j(x_{F_j})
\ud \mu}= \mathbb{E}[\zeta(X_{F})| X_{F_i}=x_{F_i}]
$$
where the function $\zeta(x_F)$ satisfies $\zeta(\{x_{it}\}_{t\in
F}) = z_i$. This step is handled by individual agents, which we have
assumed can be done perfectly. Therefore, if we concentrate on the
functional evolution level (instead of on the actual data), the
algorithm can be interpreted in terms of iterative conditional
expectations:

\begin{procedure}[H]
\SetLine $g_j(x_{F_j})\leftarrow 0, \forall j\in \{1,\cdots,D\}$\;
$\zeta(x_F) \leftarrow \phi(x_F)$\; $err_{new} \leftarrow
\mathbb{E}[\zeta^2(X_{F})]$\; $err_{old} \leftarrow 0$\;
\While{$|err_{old} - err_{new}|
> \epsilon$}{ \For{$j$ from $1$ to $D$} {
$\Delta g_j(x_{F_j}) \leftarrow \mathbb{E}[\zeta(X_{F})|
X_{F_i}=x_{F_i}]$\; $g_j(x_{F_j}) \leftarrow g_j(x_{F_j}) + \Delta
g_j(x_{F_j})$\;

$\zeta(x_F) \leftarrow \zeta(x_F) - \Delta g_j(x_{F_j})$\;
 }
 $err_{old} \leftarrow err_{new}$\;
 $err_{new} \leftarrow \mathbb{E}[\zeta^2(X_{F})]$\;
 }

\end{procedure}

Notice that the training errors are system ``biases" caused by the
limitation of our ``linear decomposition", and the effects of random
error caused by the finite number of training examples. These latter
are the same as in classical learning theory and are not factors to
be considered as resulting from the ``distributed" nature of the
problem. Thus, we do not consider them in our discussion.

From the point of view of the fusion center, it simply sends its
data that represents $\zeta(x_F)$ to an agent, waits for the agent
to first update its own estimator $g_j(x_{F_j})$, and then to send
back the difference $\Delta g_j(x_{F_j})$. The fusion center then
updates its data to represent $\zeta(x_F) - \Delta g_j(x_{F_j})$,
and moves on to the next agent.

From the point of view of an individual agent, the task is also
straightforward: when agent $j$ receives the data describing the
latest version of $\zeta(x_F)$ from the fusion center, it finds an
optimal estimator $\Delta g_j(x_{F_j})$ of $\zeta(x_F)$ based on all
the data on features in $F_j$, uses $\Delta g_j(x_{F_j})$ to update
its own estimator $g_j(x_{F_j})$, and then sends $\Delta
g_j(x_{F_j})$ back to the fusion center. In general, the algorithm
is simple, and each agent can use its own learning algorithm to
determine (approximately) the conditional-mean estimator.

It is worth noting that once the estimator is trained, it is
distributively allocated throughout the entire system. Thus, when
new data comes to the fusion center, it sends features to the
corresponding agents and then sums their estimates to form a global
estimate.

\section{Theoretical analysis of ICEA}

Intuitively, the above algorithm will repeatedly reduce the power of
the residual stored in the fusion center. But does it converge? And,
if so, what does it converge to and at what rate? Now let us look at
the answers to these questions for some special cases.

The monotonicity of the residual is easy to see. More specifically,
the root-mean-square-error (RMSE) of the ensemble estimator is
monotonically non-increasing. This is because in ICEA, we repeatedly
fix all the individual estimators but one, and optimize only that
one and use the new function to replace the old. Thus, the new
estimator cannot be worse than the old one, and therefore, the RMSE
must be non-increasing.

Moreover, since the RMSE is always non-negative, the RMSE sequence
is a monotonically non-increasing, lower bounded sequence, which
guarantees the convergence of the algorithm (if we use the change in
RMSE as the convergence criterion, which is what we did in the
algorithm previously shown). However, there is no guarantee of
uniqueness (different initial conditions might lead to different
limits), nor of equivalence between the limits and the solution to
the optimization problem given in the previous section.

So in the following subsections, we discuss the functional
convergence of ICEA under some special cases.

\subsection{Non-expansive map for two agent case}
For the two-dimensional, two-agent case, the algorithm is intended
to solve the following optimization problem:
$$
\min_{g_1,g_2} \mathbb{E}\left[\left(\phi(x_1, x_2) - g_1(x_1) -
g_2(x_2)\right)^2\right].
$$

It is straightforward to show that the optimal solution
$g_{1,opt}(x_1)$ and $g_{2,opt}(x_2)$ should satisfy equations
$$
g_1(x_1) = \mathbb{E}\left[\left(\phi(x_1, x_2) -
g_2(x_2)\right)|X_1 = x_1\right]
$$
and
$$
g_2(x_2) = \mathbb{E}\left[\left(\phi(x_1, x_2) -
g_1(x_1)\right)|X_2 = x_2\right]
$$
simultaneously.

On the other hand, if we apply ICEA to the two dimensional
distributed learning problem, we will iteratively find the solutions
to the equations above. And (hopefully) the solution will converge
to the desired $g_{1,opt}(x_1)$ and $g_{2,opt}(x_2)$; i.e. ICEA
enables us to approximate the solution to a difficult optimization
problem by solving a sequence of simplified optimization problems
iteratively. Of course, rigorously, we need to prove the convergence
of this algorithm and the uniqueness of its limit.

Ideally, if we can show that each round of the algorithm is actually
a contractive map on a well-defined metric space, it is easy to
apply the fixed point theorem to guarantee the uniqueness of the
limit.

Unfortunately, we can prove only a weaker conclusion: for the
two-agent case, ICEA, after each complete round (i.e. after each
agent updates its estimator), is equivalent to a non-expansive map.

First we need to define a suitable measure of distance between two
functions $g(x_F)$ and $h(x_F)$:
$$
d\left( g(x_F),h(x_F) \right) = \mathbb{E} \left[ (g(X_F)-h(X_F))^2
\right].
$$

The algorithm performs the following operation to a function
$g_1(x_1)$ after each complete round (denote the mapping as $T$):
\begin{eqnarray*}
& &T\{g_1(X_1)\}=\\
 & & \mathbb{E}[\phi(X_1,X_2) -
\mathbb{E}[\phi(X_1,X_2) - g_1(X_1)|X_2]|X_1=x_1].
\end{eqnarray*}
Therefore, the distance between $T\{g_1(x_1)\}$ and
$T\{g_1^{*}(x_1)\}$ is given by
\begin{displaymath}
\mathbb{E}\left[\left(\mathbb{E}[\mathbb{E}[g_1(X_1)-g_1^{*}(X_1)|X_2]|X_1]\right)^2\right].
\end{displaymath}
In order to show that $T$ is a non-expansive map, it is equivalent
to prove the following inequality:
\begin{displaymath}
\mathbb{E}\left[\left(\mathbb{E}[\mathbb{E}[g(X_1)|X_2]|X_1]\right)^2\right]\le
\mathbb{E}[g^2(X_1)],
\end{displaymath}
where $g(X_1) = g_1(X_1)-g_1^{*}(X_1)$.

Define $\mu_g = \mathbb{E}[g(X_1)]$, and notice two facts:
\begin{eqnarray*}
&
&\mathbb{E}\left[\left(\mathbb{E}[\mathbb{E}[g(X_1)|X_2]|X_1]\right)^2\right]
- \mu_g^2\\
&=&\mathbb{E}\left[\left(\mathbb{E}[\mathbb{E}[g(X_1)|X_2]|X_1]
-\mu_g\right)^2\right],
\end{eqnarray*}
and
$$
\mathbb{E}\left[g^2(X_1)\right] - \mu_g^2=
\mathbb{E}\left[\left(g(X_1)-\mu_g\right)^2\right].
$$
Then, the original inequality is equivalent to the inequality
\begin{displaymath}
\mathbb{E}\left[\left(\mathbb{E}[\mathbb{E}[g(X_1)|X_2]|X_1]
-\mu_g\right)^2\right]\le\mathbb{E}\left[\left(g(X_1)-\mu_g\right)^2\right].
\end{displaymath}
Then, we have that the left hand side satisfies
\begin{eqnarray*}
LHS &=& \mathbb{E}\left[\left(\mathbb{E}[\mathbb{E}[g(X_1)|X_2]|X_1]
-\mathbb{E}[\mathbb{E}[g(X_1)]|X_1]\right)^2\right]\\
 &=& \mathbb{E}\left[\left(\mathbb{E}[\mathbb{E}[g(X_1)|X_2]-\mathbb{E}[g(X_1)]|X_1]\right)^2\right]\\
&\le&\mathbb{E}\left[\mathbb{E}[\left(\mathbb{E}[g(X_1)|X_2]-\mathbb{E}[g(X_1)]\right)^2|X_1]\right]\\
&=&\mathbb{E}\left[\left(\mathbb{E}[g(X_1)|X_2]-\mathbb{E}[g(X_1)]\right)^2\right].
\end{eqnarray*}
The inequality step is because of Jensen's inequality:
\begin{displaymath}
\phi[\mathbb{E}(X)] \le \mathbb{E}[\phi(X)]
\end{displaymath}
when $\phi$ is a (measurable) convex function. Moreover, we also
have
$$
\mathbb{E}\left[ \left( \mathbb{E}[g(X_1)|X_2] -
\mathbb{E}[g(X_1)]\right)^2 \right]=\Var\left[ \left(
\mathbb{E}[g(X_1)|X_2]\right)^2 \right],
$$
and hence,
\begin{displaymath}
LHS \le \Var\left[ \left( \mathbb{E}[g(X_1)|X_2]\right)^2 \right].
\end{displaymath}
In addition, we have that the right hand side satisfies
\begin{displaymath}
RHS = \Var[g(X_1)],
\end{displaymath}
and an important relationship:
\begin{displaymath}
\Var[g(X_1)] \ge \Var\left[ \left( \mathbb{E}[g(X_1)|X_2]\right)^2
\right] + \mathbb{E}\left[ \Var[g(X_1)|X_2] \right].
\end{displaymath}
Therefore,
\begin{displaymath}
RHS \ge LHS + \mathbb{E}\left[ \Var[g(X_1)|X_2] \right] \ge LHS,
\end{displaymath}
and hence we have proven that $T$ is an non-expansive map. This
result can be easily generalized to the two-agent, high dimensional
case.

\subsection{Contractive map for a special case}
Non-expansiveness is weaker than contractiveness, and there is no
general ``fixed-point" theorem. But, under certain conditions, we
are able to draw stronger conclusions. For instance, if we restrict
the problem to the two-dimensional case, restrict the hidden rule to
be a finite order bivariate polynomial, and restrict the
distribution of the dependent variables to the two-dimensional joint
Gaussian distribution with correlation coefficient $|\rho|<1$, then
ICEA can be shown to be a contractive map.

Moreover, for the two-agent, two-dimension Gaussian case above, we
can also measure the speed of convergence by the contractive factor
of the contractive map. It can be shown that the factor is $\rho^4$,
i.e.\footnote{This is shown in the appendix.}
$$d\left(T(g(X_1)),T(h(X_1))\right)\le \rho^4 d\left(g(X_1), h(X_1)\right).$$
So when the two dimensions are weakly correlated, the convergence
can be very fast.

\section{Simulation results of ICEA}

\subsection{Simulation in Terms of Function Evolution}
A detailed simulation of the two-agent, two-dimensional,
finite-order bivariate-polynomial hidden rule, jointly-Gaussian case
is shown below. The hidden rule is
$$\phi(x_1,x_2) = x_1x_2^2+x_1^2+2,$$
and $(X_1, X_2)$ is jointly Gaussian with zero mean, unit variance
and correlation coefficient $\rho = 1/2$.

\begin{figure}[!hbtp] 
  \centering
  \includegraphics[scale = 0.3]{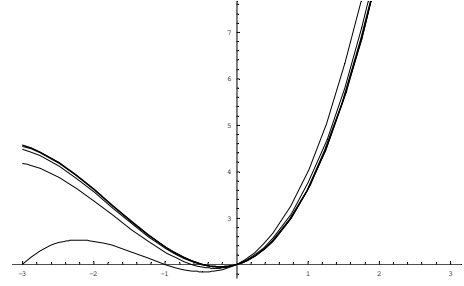}
  \caption{The convergence of $g_1(x_1)$.}
  \label{figgx}
\end{figure}

\begin{figure}[!hbtp] 
  \centering
  \includegraphics[scale = 0.3]{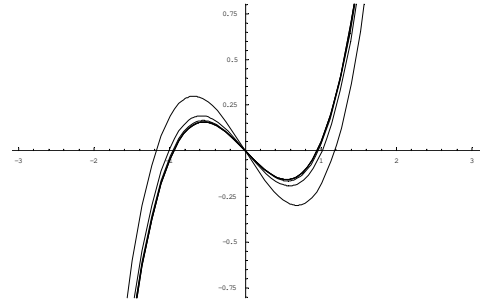}
  \caption{The convergence of $g_2(x_2)$.}
  \label{fighy}
\end{figure}

On initializing $g_1(x_1)$ and $g_2(x_2)$ to be $0$ and applying
ICEA to the problem, we get the following results shown in Table
\ref{tab1}:

\begin{table}[!hbtp]
\centering
\begin{tabular}{c|c|c}
  \hline
  Round & $g_1(x_1)$ and $g_2(x_2)$ & RMSE\\
  \hline
  1 & $2+.7500x_1+x_1^2+.2500x_1^3$ & $ $\\ \cline{2-2}
    & $-.6563x_2+.4688x_2^3$ & $1.4941406250$\\
  \hline
  2 & $2+.5508x_1+x_1^2+.1914x_1^3$ & $ $\\ \cline{2-2}
    & $-.4907x_2+.4761x_2^3$ & $1.2974381447$\\
  \hline
  3 & $2+.4598x_1+x_1^2+.1905x_1^3$ & $ $\\ \cline{2-2}
    & $-.4442x_2+.4762x_2^3$ & $1.2864467088$\\
  \hline
  4 & $2+.4364x_1+x_1^2+.1905x_1^3$ & $ $\\ \cline{2-2}
    & $-.4325x_2+.4762x_2^3$ & $1.2857600621$\\
  \hline
  5 & $2+.4305x_1+x_1^2+.1905x_1^3$ & $ $\\ \cline{2-2}
    & $-.4295x_2+.4762x_2^3$ & $1.2857171467$\\
  \hline
  6 & $2+.4291x_1+x_1^2+.1905x_1^3$ & $ $\\ \cline{2-2}
    & $-.4288x_2+.4762x_2^3$ & $1.2857144645$\\
  \hline
  Limit & $2+3/7~x_1+x_1^2+4/21~x_1^3$ & $ $\\ \cline{2-2}
    & $-3/7~x_2+10/21~x_2^3$ & $9/7$\\
  \hline
\end{tabular}
  \caption{Step-by-step results of the ICEA.}
  \label{tab1}
\end{table}
\noindent where the limit function is the unique solution to
equations
\begin{eqnarray*}
g_1(x_1) &=& \int_{-\infty}^{\infty}\left( \phi(x_1,x_2) - g_2(x_2)\right) f_{X_2|X_1}(x_2|x_1) \ud x_2\\
g_2(x_2) &=& \int_{-\infty}^{\infty}\left( \phi(x_1,x_2) -
g_1(x_1)\right) f_{X_1|X_2}(x_1|x_2) \ud x_2.
\end{eqnarray*}
The evolution of the functions of Table \ref{tab1} is shown in
Fig.\ref{figgx} and Fig.\ref{fighy}. It is quite clear that there is
no visible difference after a few rounds of iterations.

\begin{figure}[!hbtp] 
  \centering
  \includegraphics[scale = 0.35]{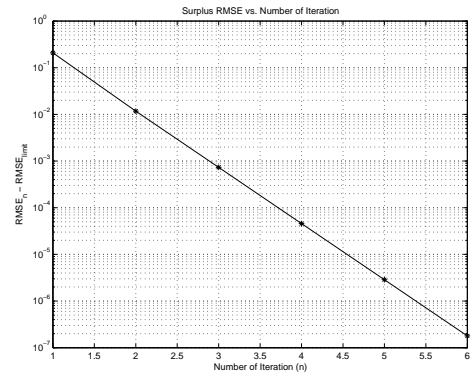}
  \caption{The rate of RMSE convergence for the two-agent, two-dimensional, joint Gaussian case.}
  \label{RMSE}
\end{figure}

Moreover, the speed of convergence can be measured (approximately)
by the surplus RMSE (the difference between the RMSE of the ensemble
estimator after the $n$th iteration and the limit RMSE) as shown in
Fig.\ref{RMSE}. Also notice that in the semi-logarithm plot, the
slope $k$ of the line is $-2.79375$, and $(e^k)^{1/4} \approx 0.5 =
\rho$, which is compatible with our theory in the previous section.

\subsection{Simulation on Real Data}
Our discussion above always assumes enough training data and perfect
individual agents that can find the MMSE estimator. However, to
justify the efficacy of ICEA in solving real problems, we test the
algorithm with real data, contrary to the functional ``simulation"
we did in the previous section.

In order to compare the distributed regression to other
multi-dimensional regression algorithms, we use three functions used
in \cite{ref8} (originally from \cite{refa1} and \cite{refa2}) as
the hidden rule to generate our simulation training data sets. The
three functions are
\begin{itemize}
\item Friedman-1: $$\phi(\mathbf{x}) = 10\sin(\pi x_1 x_2)+20(x_3-1/2)^2+10x_4+5x_5+w$$ where
$x_j \sim U[0,1], j=1\ldots5$;
\item Friedman-2: $$\phi(\mathbf{x}) = \left( x_1^2+\left( x_2x_3-\frac{1}{x_2x_4} \right)^2
\right)^{\frac{1}{2}}+w$$ where $$x_1\sim U[1,100], $$ $$x_2\sim
U[40\pi, 560\pi],$$ $$ x_3, x_5\sim U[0,1],$$ $$ x_4\sim U[1,11].$$
\item Friedman-3: $$\phi(\mathbf{x}) = \tan^{-1}
\left(\frac{x_2x_3-\frac{1}{x_2x_4}}{x_1}\right)+w$$ where the
distribution of the features are the same as that of Friedman-2.
\end{itemize}

All the feature variables are independent, and before running the
algorithm, the outcomes are normalized to the range $[0,1]$. Also,
to highlight the effect of distributiveness of the system, the
independent white noise $w$ is set to zero in our simulation. Also,
it is worth pointing out that in Friedman-2 and Friedman-3, feature
$X_5$ is irrelevant, and is set up as a test of the algorithm's
resistance to irrelevant features.

Moreover, we did the experiments on three different types of
distributed regression systems:
\begin{itemize}
\item System-1: five 1-dimensional agents
$$\{X_1\}, \{X_2\}, \{X_3\}, \{X_4\}, \{X_5\}$$
\item System-2: three 2-dimensional agents $$\{X_1, X_2\}, \{X_2, X_3\}, \{X_4,
X_5\}$$
\item System-3: two 5-dimensional agents $$\{X_1, X_2, X_3\}, \{X_2, X_4, X_5\}$$
\end{itemize}
In all these cases, all the features are fully covered by the union
of the features observable by individual agents.

We use a regression tree - a commonly used ``weak learner" in the
boosting algorithms - as our learning algorithm for the individual
agents. Notice that $L_2$-regression boosting (introduced in
\cite{ref10}) is equivalent to a one-agent system, in which all the
dimensions are accessible by the agent. In this sense,
$L_2$-regression boosting is the ``limit algorithm" of the ICEA for
distributed systems. So it is natural to compare the performance of
distributed system to $L_2$-regression boosting. Also, to compare
the cooperative algorithm ICEA to non-cooperative algorithm (like
the algorithms in \cite{ref2}), we also ran the data on a
hierarchical algorithm. The individual agents are identical to those
of ICEA, and the fusion center can further train an estimator using
$L_2$-regression boosting, taking the output of the agents as the
features.

\begin{figure}[!hbtp] 
  \centering
  \includegraphics[scale = 0.5]{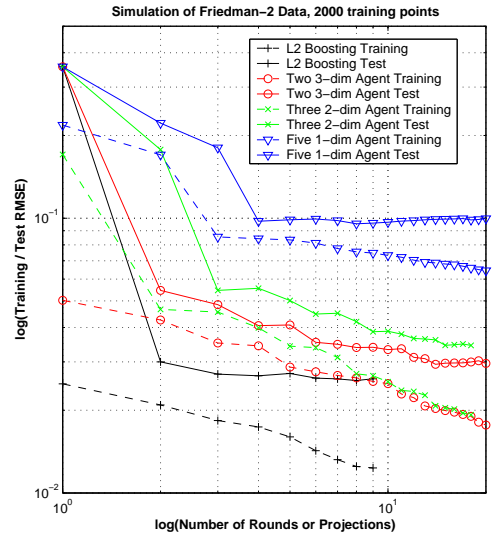}
  \caption{Simulation using Friedman-2 data with 2000 training data points and 4000 test data points. The training error and test error for $L_2$-regression boosting and ICEA with three different distributed systems: five 1-dim agents, three 2-dim agents and two 3-dim agents. The training error (dashed lines) and test error (solid lines) of
ICEA decreases monotonically, and converges quite fast. And systems
with high-dimensional individual agents have lower training error
and test error than systems with low-dimensional individual agents.}
  \label{MCSimu}
\end{figure}

Running three different types of algorithms on three different data
sets, with 2000 training data points and 4000 test data points, the
mean squared errors are shown in Table \ref{tab2}, and the detailed
plot of training/test error after each round of $L_2$-regression
boosting and ICEA with three different distributed systems running
on data set Friedman-2 are shown in Fig. \ref{MCSimu}.

\begin{table}[!hbtp]
\centering
\begin{tabular}{c|c|c|c|c}
  \hline
  Data Set & System & ICEA & Hierarchical & $L_2$ Boosting\\
  \hline
              &1& $.0050$ & $ .024 $ & \\ \cline{2-3}
     Friedman-1&2& $.0014$ & $.061$ & .0051 \\ \cline{2-3}
             & 3 & $.0039$ & $.036$ & \\
  \hline
                & 1 & $.010$ & $ .075 $ & \\ \cline{2-3}
     Friedman-2 & 2 & $.0012$ & $.079$ & .00066\\ \cline{2-3}
                & 3 & $.00088$ & $.13$ & \\
  \hline
             & 1 & $.0082$ & $ .35 $ & \\ \cline{2-3}
     Friedman-3&2& $.0062$ & $.23$ & .0034 \\ \cline{2-3}
             & 3 & $.0035$ & $.31$ & \\
  \hline
\end{tabular}
  \caption{Test error (mean squared error) of $L_2$-regression boosting, ICEA and hierarchical on data sets Friedman-1,-2 and -3.}
  \label{tab2}
\end{table}

As expected, $L_2$-regression boosting performs best for most of the
cases, except for Friedman-1, the hidden hypothesis of which is
basically additive. Because System-2 is not so complicated yet is
complex enough to fully capture the model, the ICEA algorithm
running on System-2 performs best. However, for Friedman-2 and
Friedman-3, where the hidden models are no longer additive,
$L_2$-regression boosting, with full access to all the dimensions,
outperforms other algorithms. And for ICEA, the performance is
better when the individual agents have access to more dimensions,
capable of describing more complex coupling among the features. The
hierarchical algorithm works (thought not so well) for additive
models, yet the algorithm performs poorly for data sets with
complicated functions where there is strong coupling among the
features. Since the estimators used for individual agents in ICEA
and the hierarchical algorithm are identical (regression trees), the
performance difference can be attributed to the benefit of applying
cooperative training in ICEA.

\section{Problems of ICEA and Possible Improvements}
\subsection{Limitations of ICEA} Of course, since we restrict our
approximation of $\phi(x_F)$ to the sum $\sum_{j=1}^D g_j(x_{F_j})$,
we lose richness of the ensemble estimator. For instance, for the
function $\phi(x_1,x_2) = x_1 x_2$ with $X_1, X_2$ being independent
standard Gaussian variables, if agent 1 has access only to dimension
1, and agent 2 has access only to dimension 2, then the linear model
estimator is simply $0$, which means nothing can be learned. So
restricting to a linear form can lead to some serious problems.
However, there are several ways to solve this problem and greatly
expand the efficacy of ICEA.

First, we can linearize the function $\phi(x_F)$. For instance, in
the example above, if we take the logarithm of the function
$\phi(x_1,x_2) = x_1 x_2$, then we get $\log(\phi(x_1,x_2)) =
\log(x_1)+ \log(x_2)$. In this case, we can use the linear additive
model to accurately depict the ensemble function. Therefore, with a
proper non-linear transformation, we can greatly expand the scope of
problems that can be optimally solved by our linear additive model.

Second, we can project the function $\phi(x_F)$ on more linear
combinations of the features of the agents. For instance, in the
above problem, if we have two other agents that have access to data
$x_1 + x_2$ and $x_1 - x_2$, then the model can also be accurately
learned by these two agents. In practice, because there is
significant redundancy amongst the data of the agents, we don't need
to intentionally calculate these linear combinations (which requires
more communication resources). Instead, we simply take advantage of
the redundancies contained in the data, which is often considered to
be a hazard in some learning algorithms.

\subsection{Developing More Intelligent Algorithms}
Boosting for regression sheds light on the design of algorithms more
``intelligent" than ICEA, which simply refits the residual on each
agent one after another. ICEA can be improved in more ways than one,
in terms of increasing speed of convergence, finding a natural
stopping rule to avoid over-training and to reduce generalization
error. Several rudimentary experiments have shown that, instead of
iteratively refitting the residual one agent after another to reduce
merely the training error, we are better off if we choose among the
agents more intelligently, and take both the training error on the
residual and the complexity of the model into consideration. For
instance, a greedy algorithm that always chooses the agent that
provides the minimum training error can greatly increase the speed
of convergence, and an algorithm using the size of the decision tree
as a penalty term can effectively reduce overtraining. It is
worthwhile to explore more subtle rules of selecting estimators from
agents and more delicate ways to combine them. ICEA is perhaps the
most intuitive algorithm, but far from the optimal one.

\section{Conclusions}
By restricting the ensemble estimator to an additive form (linear
combination of individual estimators), we have developed an
iterative algorithm (ICEA) that is guaranteed, under certain
conditions, to converges to a unique limit function (or rule, or
hypothesis). This limit is an approximation of the true function,
and, with the help of some additional features (linearization,
redundant data), the approximation can be quite accurate. ICEA also
works quite well with real data, with enough training points and
properly selected individual estimators. By sending only the
predictions and withhold the data, ICEA also preserves the privacy
of data of individual agents. There are still many aspects of the
algorithm that can be changed to improve the performance of
distributed regression, and these are of interest for further
investigation.

\section{Appendix}
\emph{Lemma 1} Suppose $g$ is a polynomial of order $M$,
$$g(x)=\sum_{n=0}^M a_n x^n,$$
and $g'(x)$ is given by
$$g'(x) = \int_{-\infty}^{\infty} \left( \int_{-\infty}^{\infty} g(x) f_{X|Y}(x|y)\ud x\right) f_{Y|X}(y|x) \ud
y.$$ Then, with the additional assumption that
$$\int_{-\infty}^{\infty}  g(x)f_X(x) \ud x = 0.$$
we have inequality
$$\frac{\int_{-\infty}^{\infty} g'^2(x)f_X(x) \ud x}{\int_{-\infty}^{\infty} g^2(x)f_X(x) \ud x} \le
\rho^4,$$ where $f_{X|Y}$, $f_{Y|X}$, and $f_X$ are all probability
densities derived from the joint Gaussian distribution of zero mean,
unit variance and correlation $\rho$.

\emph{Proof}: The conditional distribution of $X$ given $Y$ and the
distribution of $Y$ given $X$ are
$$ X|Y \sim N(\rho y, 1-\rho^2),$$
$$ Y|X \sim N(\rho x, 1-\rho^2).$$
Therefore, we have
$$g'(x)=\sum_{n=0}^M a_n \frac{\ud^n}{\ud
t^n}\exp{\{ \rho^2 x t + \frac{(1-\rho^4)t^2}{2} \}
}\arrowvert_{t=0}.$$ Notice that the exponential term, with proper
manipulation, can be expressed in the form of the sum of Hermitian
polynomials. Thus, on defining
$$X=\frac{\rho^2}{i\sqrt{2(1-\rho^2)}}x,s = i\sqrt{\frac{1-\rho^4}{2}}t,$$
we have $$\exp{\{ \rho^2 x t + \frac{1-\rho^4}{2}t^2\}} =
e^{2Xs-s^2} = \sum_{k=0}^\infty H_k(X)\frac{s^k}{k!}.$$ Then the
expression of $g'(x)$ can be rewritten as
\begin{eqnarray*}
g'(x) &=& \sum_{n=0}^M a_n\frac{\ud^n}{\ud
t^n}\left(\sum_{k=0}^\infty H_k(X)\frac{s^k}{k!}\right)|_{s=0}\\
&=& \sum_{n=0}^M a_n
H_n(X)\left(i\sqrt{\frac{1-\rho^4}{2}}\right)^n.
\end{eqnarray*}
Therefore, we have a closed-form expression for $g'(x)$, which is
also an $M^\mathrm{th}$-order polynomial:
\begin{eqnarray*}
g'(x) &=&\sum_{n=0}^M a_n
H_n\left(\frac{\rho^2}{i\sqrt{2(1-\rho^4)}}x\right)\left(i\sqrt{\frac{1-\rho^4}{2}}\right)^n\\
&=& \sum_{n=0}^M a_n
\sum_{k=0}^{\left[\frac{n}{2}\right]}\frac{n!}{k!(n-2k)!}\left(\frac{1-\rho^4}{2}\right)^k(\rho^2x)^{n-2k}.
\end{eqnarray*}

By computation, we can derive
$$
\int_{-\infty}^{\infty}  g^2(x)f_X(x) \ud x = \sum_{n=0}^M n!\left(
\sum_{k=0}^{\left[ \frac{M-n}{2} \right]} a_{n+2k}
\frac{(n+2k)!}{2^k n! k!}\right)^2,
$$
and
$$
\int_{-\infty}^{\infty}  g'^2(x)f_X(x) \ud x  = \sum_{n=0}^M
(\rho^4)^n n!\left( \sum_{k=0}^{\left[ \frac{M-n}{2} \right]}
a_{n+2k} \frac{(n+2k)!}{2^k n! k!}\right)^2.
$$
Moreover, since $g$ is of zero mean,
$$\sum_{k=0}^{\left[\frac{M}{2}\right]}a_{2k}\frac{(2k)!}{2^k k!} = 0.$$
Therefore,
\begin{eqnarray*}
 & & \frac{\int_{-\infty}^{\infty} g'^2(x)f_X(x) \ud
x}{\int_{-\infty}^{\infty} g^2(x)f_X(x) \ud x}\\ &=&
\frac{\sum_{n=1}^M (\rho^4)^n n!\left( \sum_{k=0}^{\left[
\frac{M-n}{2} \right]}
a_{n+2k} \frac{(n+2k)!}{2^k n! k!}\right)^2}{\sum_{n=1}^M n!\left( \sum_{k=0}^{\left[ \frac{M-n}{2} \right]} a_{n+2k} \frac{(n+2k)!}{2^k n! k!}\right)^2}\\
&\le& \frac{\sum_{n=1}^M \rho^4 n!\left( \sum_{k=0}^{\left[
\frac{M-n}{2} \right]}
a_{n+2k} \frac{(n+2k)!}{2^k n! k!}\right)^2}{\sum_{n=1}^M n!\left( \sum_{k=0}^{\left[ \frac{M-n}{2} \right]} a_{n+2k} \frac{(n+2k)!}{2^k n! k!}\right)^2}\\
&=& \rho^4.\oblong
\end{eqnarray*}

If the hidden rule $\phi(x_1, x_2)$ is restricted to a bivariate
polynomial with finite order $M$ and zero mean, and $g_1(x_1),
g_2(x_2)$ are both initialized as 0, then after each iteration,
$g_1(x_1)$ and $g_2(x_2)$ will remain in the space of zero-mean
polynomials of finite order $M$. If we define the distance between
two polynomials $g_1(x)$ and $g_2(x)$ as $\int_{-\infty}^{\infty}
\left(g_1(x)-g_2(x)\right)^2 f_X(x) \ud x,$ then, according to
\emph{Lemma 1}, the map $T$ that convert $g(x)$ to $g'(x)$ is a
contractive map. Because under our definition of the distance, the
space of zero-mean polynomials is complete, we can apply the
contractive mapping theorem to guarantee the functional convergence
of ICEA for the two-agent, two-dimensional, joint-Gaussian,
finite-order-polynomial case. Moreover, also by the lemma, the
``contractive factor" of the map is $\rho^4$. As for the hidden rule
with non-zero mean, the bias is also addressed by the constant term
of the first agent, which will remain the same in the following
iterations and hence has no influence on the convergence.

%

%


\bibliographystyle{IEEEtran}

\end{document}